\input harvmac

\def\dell{\delta_\ell}

\def\epsp{\varepsilon_+}
\def\epsm{\varepsilon_-}

\def\gsim{{~\raise.15em\hbox{$>$}\kern-.85em
          \lower.35em\hbox{$\sim$}~}}
\def\lsim{{~\raise.15em\hbox{$<$}\kern-.85em
          \lower.35em\hbox{$\sim$}~}}
\def\Re{{\cal R}e}
 
\def\epst{\epsilon_\tau}

\def\epson{\epsilon_{12}}         
\def\epsonu{\epsilon^\nu_{12}}   

\noblackbox
\baselineskip 14pt plus 2pt minus 2pt
\Title{\vbox{\baselineskip12pt
\hbox{hep-ph/0002268}
\hbox{IASSNS--HEP--00--08}
}}
{\vbox{
\centerline{Pseudo-Dirac Solar Neutrinos}
  }}
\centerline{Yosef Nir}
\medskip
\centerline{\it School of Natural Science,
Institute for Advanced Study}
\centerline{\it Princeton, NJ 08540, USA\foot{Address for
academic year 1999-2000}}
\centerline{nir@ias.edu}
\medskip
\centerline{\it Department of Particle Physics}
\centerline{\it Weizmann Institute of Science, Rehovot 76100, Israel}
\centerline{ftnir@wicc.weizmann.ac.il}
\bigskip

\baselineskip 18pt
\noindent
Three of the viable solutions of the solar neutrino problem are consistent
with close to maximal leptonic mixing: $\sin^2\theta_{12}={1\over2}(1-\epson)$
with $|\epson|\ll1$. Flavor models can naturally explain close to maximal mixing
if approximate horizontal symmetries force a pseudo-Dirac structure on the neutrino 
mass matrix. An experimental determination of $|\epson|$ and sign($\epson$)
can constrain the structure of the lepton mass matrices and consequently 
provide stringent tests of such flavor models. If both $|\epson|$ and
$\Delta m^2_{21}$ are known, it may be possible to estimate the mass scale of 
the pseudo-Dirac neutrinos. Radiative corrections to close to maximal mixing
are negligible. Subtleties related to the kinetic terms in Froggatt-Nielsen
models are clarified.  

\Date{2/00}

%%%%%%%%%%%%%%
\newsec{Introduction}
\nref\BKS{J.N. Bahcall, P.I. Krastev and A.Yu. Smirnov,
 Phys. Rev. D60 (1999) 093001, hep-ph/9905220;
 Phys. Lett. B477 (2000) 401, hep-ph/9911248.}%
\nref\GHPV{M.C. Gonzalez-Garcia, P.C. de Holanda, C. Pena-Garay and
 J.W.F. Valle, Nucl. Phys. B573 (2000) 3, hep-ph/9906469.}% 
\nref\FLMP{G.L. Fogli, E. Lisi, D. Montanino and A. Palazzo,
 Phys. Rev. D62 (2000) 013002, hep-ph/9912231.}%
\nref\GGG{C. Giunti, M.C. Gonzalez-Garcia and C. Pena-Garay,
 Phys. Rev. D62 (2000) 013005, hep-ph/0001101.}% 
Three of the solutions of the solar neutrino problem require
a large mixing angle \refs{\BKS-\GGG}:
\eqn\SNdata{\eqalign{
{\rm LMA}:&\ \ \ \sin^22\theta_{12}\sim0.7-1,\ \ \ 
\Delta m^2_{21}\sim(1-20)\times10^{-5}\ eV^2,\cr
{\rm LOW}:&\ \ \ \sin^22\theta_{12}\sim0.8-1,\ \ \ 
\Delta m^2_{21}\sim(3-30)\times10^{-8}\ eV^2,\cr
{\rm VAC_{\rm L}}:&\ \ \ \sin^22\theta_{12}\sim0.7-1,\ \ \ 
\Delta m^2_{21}\sim(4-10)\times10^{-10}\ eV^2.\cr}}
Here LMA and LOW refer to matter-enhanced oscillations with a large
mixing angle in the high and low $\Delta m^2$ ranges, respectively,
while VAC$_{\rm L}$ refers to vacuum oscillations with relatively 
large $\Delta m^2$.
The range for the mixing angle in \SNdata\ is close to maximal mixing,
$\sin^22\theta_{12}=1$. This case is particularly interesting from the 
theoretical point of view. It follows from a simple structure of the 
relevant $2\times2$ block in the neutrino mass matrix in the basis where 
the charged lepton mass matrix is diagonal:
\eqn\Mnumax{M_\nu^{(2)}=m\pmatrix{0&1\cr 1&0\cr}.}
Such a structure is easily obtained in models of horizontal symmetries
\nref\Petc{S.T. Petcov, Phys. Lett. B110 (1982) 245.}%
\nref\LePe{C.N. Leung and S.T. Petcov, Phys. Lett. B125 (1983) 461.}%
\nref\BGL{G.C. Branco, W. Grimus and L. Lavoura,
 Nucl. Phys. B312 (1989) 492.}%
\nref\PeSm{S.T. Petcov and A.Yu. Smirnov,
 Phys. Lett. B322 (1994) 109, hep-ph/9311204.}%
\nref\BLPR{P. Binetruy, S. Lavignac, S. Petcov and P. Ramond,
 Nucl. Phys. B496 (1997) 3, hep-ph/9610481.}%
\refs{\Petc-\BLPR}\ that try to explain the observed smallness and hierarchy 
in the charged fermion parameters (mass ratios and mixing angles). 
For example, if the lepton doublets of the first two generations carry an 
opposite charge under a U(1) symmetry (and the relevant scalar field is
neutral), then $M_\nu^{(2)}$ has the structure \Mnumax\ in the symmetry limit.

\nref\LNSa{M. Leurer, Y. Nir and N. Seiberg,
 Nucl. Phys. B398 (1993) 319, hep-ph/9212278.}%
Any horizontal symmetry must be broken in Nature. An unbroken horizontal
symmetry leads to either degeneracy between fermions of different generations
or vanishing mixing angles (see {\it e.g.} \LNSa\ and references therein).
In particular, the mass degeneracy implied by \Mnumax\ must be broken
to satisfy \SNdata. The horizontal symmetry still has observable consequences 
if the breaking parameters are small. 
Then the low energy effective theory is subject to selection rules that are 
manifested in the smallness and hierarchy of the flavor parameters. In the 
case of close-to-maximal mixing, the small breaking leads to a small splitting 
between the masses of the two neutrinos and to a small deviation from 
maximal mixing, that is, the two Majorana neutrinos form
a {\it pseudo-Dirac} neutrino:
\eqn\cltoma{{\Delta m^2_{21}\over m^2}\ll1,\ \ \ \ 1-\sin^22\theta_{12}\ll1.}
Here $m$ denotes the average of $m_1$ and $m_2$.
A measurement of these small effects will provide further information 
about the pattern of symmetry breaking and guide us in the process of
selecting among the many presently viable models of horizontal symmetries. 
(For interesting studies of the implications of solar neutrino
measurements for small entries in the neutrino mass matrix, see
\nref\Akh{E.Kh. Akhmedov, Phys. Lett. B467 (1999) 95, hep-ph/9909217.}%
\nref\ABR{E.Kh. Akhmedov, G.C. Branco and M.N. Rebelo,
 Phys. Rev. Lett. 84 (2000) 3535, hep-ph/9912205.}%
refs. \refs{\Akh,\ABR}.)

There are three light active neutrinos in Nature. (In this work we assume
that these are the only light neutrinos and do not
consider the possibility of light sterile neutrinos. Note that the large 
angle solutions of the solar neutrino problem are inconsistent with a 
pseudo-Dirac $\nu_e-\nu_s$ combination, such as in the model of ref.
\ref\SiVa{J.W. Valle and M. Singer, Phys. Rev. D28 (1983) 540.}.) 
In the two generation framework, where there is a single mixing
angle $\theta$, maximal mixing is defined by maximal oscillation 
depth {\it in vacuum} and corresponds to $\sin^22\theta=1$.
In the three generation framework, what we mean by maximal mixing 
is that the disappearance probability is equivalent 
to that for maximal two neutrino mixing at the relevant mass scale
 \ref\BPWW{V. Barger, S. Pakvasa, T.J. Weiler and K. Whisnant,
 Phys. Lett. B437 (1998) 107, hep-ph/9806387.}.
The disappearance probability for $\nu_e$ in vacuum,
$P_{e\not e}$, is given by
\eqn\Penote{P_{e\not e}=4|V_{e1}|^2|V_{e2}|^2\sin^2\Delta_{\rm sun}
-4|V_{e3}|^2\sin^2\Delta_{\rm atm}.}
Here $V_{ij}$ are the elements of the Maki-Nakagawa-Sakata (MNS) matrix
\ref\MNS{Z. Maki, M. Nakagawa and S. Sakata,
 Prog. Theo. Phys. 28 (1962) 870.}\
and we use the definition $\Delta_{jk}\equiv {\Delta m^2_{jk} L\over4E}$,
and the following input from solar and atmospheric neutrino experiments:
\eqn\hiedms{|\Delta_{\rm sun}|=|\Delta_{21}|\ll
|\Delta_{31}|\simeq|\Delta_{32}|=|\Delta_{\rm atm}|.}
At the $L/E$-scale that is relevant to solar neutrinos, the second
term in \Penote\ averages out to $-2|V_{e3}|^2$. The only oscillatory 
term is the first one, and our definition of maximal mixing corresponds to
\eqn\maxthr{4|V_{e1}|^2|V_{e2}|^2=1,}
which leads to
\eqn\maxth{|V_{e1}|^2=|V_{e2}|^2=1/2,\ \ \ \ |V_{e3}|^2=0.}
In the standard parametrization of the $V_{\rm MNS}$ matrix,
\eqn\stapar{V_{\rm MNS}=\pmatrix{
c_{12}c_{13}&s_{12}c_{13}&s_{13}e^{-i\delta}\cr
-s_{12}c_{23}-c_{12}s_{23}s_{13}e^{i\delta}&
c_{12}c_{23}-s_{12}s_{23}s_{13}e^{i\delta}& s_{23}c_{13}\cr
s_{12}s_{23}-c_{12}c_{23}s_{13}e^{i\delta}&
-c_{12}s_{23}-s_{12}c_{23}s_{13}e^{i\delta}& c_{23}c_{13}\cr},}
where $s_{ij}\equiv\sin\theta_{ij}$ and $c_{ij}\equiv\cos\theta_{ij}$,
the conditions \maxth\ translate into
\eqn\maxpar{s^2_{12}=1/2,\ \ \ \ s_{13}=0.}
By {\it close-to-maximal-mixing} we refer to a situation close to 
\maxth\ or, equivalently, to \maxpar:
\eqn\ctmpar{\epson\equiv 1-2s_{12}^2\ll1,\ \ \ s_{13}\ll1.}
Our convention here is that $\Delta m^2_{21}\equiv m_2^2-m_1^2>0$,
so that $\epson>0$ ($\epson<0$) corresponds to a situation where
the lighter (heavier) state has a larger component of $\nu_e$.

\nref\GRS{A.H. Guth, L. Randall and M. Serna,
 JHEP 9908 (1999) 018, hep-ph/9903464.}% 
\nref\GFM{A. de Gouvea, A. Friedland and H. Murayama, hep-ph/9910286;
 hep-ph/0002064.}%
\nref\Frie{A. Friedland, hep-ph/0002063.}%
Solar neutrino experiments (and, more generally, any oscillation experiments)
are sensitive to the mass-squared difference $\Delta m^2_{12}$ but not to
the masses themselves. On the other hand, they can be sensitive to small
deviations from maximal mixing \refs{\GRS-\Frie}. Moreover, matter oscillations 
(but not vacuum oscillations) are affected differently by $\epson>0$ and 
by $\epson<0$, that is, they are sensitive not only to $\sin^22\theta_{12}$ but 
also to $\sin^2\theta_{12}$. In other words, if the solar neutrino problem
is solved by one of the large angle solutions, then experiments may provide 
us with a measurement of the sign and the size of the small parameter 
$\epson$. The purpose of this work is to understand the potential lessons for 
flavor model building from solar neutrino measurements of $\epson$.

Our interest lies in models where $\epson$, $s_{13}$ and $\Delta m^2_{21}/m^2$
are naturally small. We focus on models where there are
no exact relations between entries of the lepton mass matrices
(beyond the symmetric structure of the neutrino Majorana mass matrix).
The smallness of physical parameters must then be related to
the smallness of various entries in the mass matrices and not to
fine tuned cancellations between various contributions. 
As a concerete example of such a framework we think of models of
approximate Abelian horizontal symmetries, but most of our results
have more general applicability.

Horizontal symmetries constrain the structure of the mass matrices
$M_\nu$ and $M_\ell$. In section 2 we derive the dependence of the
mixing angles and of $\Delta m^2_{12}/m^2$ on the entries of the lepton 
mass matrices. Usually, the constraints of the horizontal symmetries apply at
a high energy scale. The effects of renormalization group evolution (RGE)
are analyzed in section 3. For specific high energy theories of flavor,
such as the Froggatt-Nielsen mechanism
\ref\FrNi{C.D. Froggatt and H.B. Nielsen, Nucl. Phys. B147 (1979) 277.},
the kinetic terms are corrected in a flavor dependent way when heavy
degrees of freedom are integrated out. The effects of non-canonical
kinetic terms are studied in section 4. The analysis in sections 2$-$4
is carried out under the simplifying assumption of CP symmetry.
Effects of phases are studied within a two generation model in section 5.
We apply our results to various models of Abelian flavor symmetries in section 6.
We summarize our conclusions in section 7.

%%%%%%%%%%%%%%%%%%%%%%
\newsec{From Interaction Basis Parameters to Physical Parameters}
Flavor models and, in particular, models with horizontal symmetries,
constrain the entries of the lepton mass matrices in the interaction
basis. To understand the implications of experimental constraints,
one needs to express the physical observables (masses and mixing
angles) in terms of the interaction basis parameters.

Given the charged lepton mass matrix $M_\ell$
and the neutrino mass matrix $M_\nu$ in some interaction basis, 
\eqn\Lmass{-{\cal L}_M=\pmatrix{{\overline{e_L}}&
{\overline{\mu_L}}&{\overline{\tau_L}}\cr}M_\ell
\pmatrix{e_R\cr\mu_R\cr\tau_R\cr}+
\pmatrix{\nu_e^T&\nu_\mu^T&\nu_\tau^T\cr}M_\nu
\pmatrix{\nu_e\cr\nu_\mu\cr\nu_\tau\cr}+{\rm h.c.},}
$V_{\rm MNS}$ can be found from the diagonalizing matrices 
$V_{\ell}$ and $V_\nu$:
\eqn\VVVP{V_{\rm MNS}=P_\ell V_{\ell}V_\nu^\dagger,}
where $P_\ell$ is a diagonal phase matrix. The unitary matrices
$V_{\ell L}$ and $V_\nu$ are found from
\eqn\diagon{V_{\ell L} M_\ell M_\ell^\dagger V_{\ell L}^\dagger=
{\rm diag}(m_e^2,m_\mu^2,m_\tau^2),\ \ \  
V_{\nu} M_\nu^\dagger M_\nu V_{\nu}^\dagger=
{\rm diag}(m_1^2,m_2^2,m_3^2).}

Our first step is to express the physical mixing angles in terms
of the parameters of the diagonalizing matrices. For simplicity,
we ignore CP violation, so that the mass matrices and, consequently, 
the diagonalizing matrices are real. (We comment on the effects
of CP violating phases in section 5.) Let us define the three unitary 
matrices
\eqn\defRmat{\eqalign{
R_{12}(\theta_{12})\ \equiv&\ \pmatrix{c_{12}&s_{12}&0\cr
-s_{12}&c_{12}&0\cr 0&0&1\cr},\cr 
R_{13}(\theta_{13})\ \equiv&\ \pmatrix{c_{13}&0&s_{13}\cr
0&1&0\cr -s_{13}&0&c_{13}\cr},\cr
R_{23}(\theta_{23})\ \equiv&\ \pmatrix{1&0&0\cr
0&c_{23}&s_{23}\cr 0&-s_{23}&c_{23}\cr}.\cr}}
Then, eq. \stapar\ (with $\delta$ set to zero) can be rewritten as
\eqn\parmns{V_{\rm MNS}\ =\ R_{23}(\theta_{23})R_{13}(\theta_{13})
R_{12}(\theta_{12}).}
We further parametrize the diagonalizing matrices as follows:
\eqn\parvnuell{\eqalign{
V_\nu^\dagger\ =&\ R_{23}(\theta_{23}^\nu)R_{13}(\theta_{13}^\nu)
R_{12}(\theta_{12}^\nu),\cr
V_\ell\ =&\ R_{23}(-\theta_{23}^\ell)R_{13}(-\theta_{13}^\ell)
R_{12}(-\theta_{12}^\ell).\cr}}

We limit ourselves to the large class of models where there are no 
{\it exact} relations between the entries in $M_\nu$ (up to the fact 
that it is symmetric, that is, $(M_\nu)_{ij}=(M_\nu)_{ji}$) and in $M_\ell$.
Then the smallness of $\epson$ and $s_{13}$ requires that the following
parameters are small:
\eqn\smallpar{s_{12}^\ell,\ \ s_{13}^\ell,\ \ \epsonu,\ \ s_{13}^\nu\ \ll1,}
where
\eqn\defepson{\epsonu\equiv1-2(s_{12}^\nu)^2.}
Evaluating to first order in the small parameters of \smallpar, we obtain:
\eqn\mnssma{\eqalign{
\epson\ =&\ \epsonu+2c_{23}^\nu s_{12}^\ell-2s_{23}^\nu s_{13}^\ell,\cr
s_{13}\ =&\ s_{13}^\nu-s_{23}^\nu s_{12}^\ell -c_{23}^\nu s_{13}^\ell.\cr}}
We caution the reader that the sign of the terms that depend on
$s_{12}^\ell$ and $s_{13}^\ell$ is ambiguous. In particular, we approximated 
$\sin2\theta_{12}^\nu=1$, but with the parametrization \parvnuell\ it could equal
$-1$. A full treatment of the sign and phase dependence of the $s_{12}^\ell$
contribution to $\epson$ is given in section 5.

\nref\HaRa{L.J. Hall and A. Rasin, Phys. Lett. B315 (1993) 164.}%
\nref\LNSb{M. Leurer, Y. Nir and N. Seiberg,
 Nucl. Phys. B420 (1994) 468, hep-ph/9410320.}%
Our next step is to express the parameters of the diagonalizing
matrices in terms of the mass matrices. For the charged lepton sector,
the expressions can be found in refs. \refs{\HaRa,\LNSb}. Typically,
one finds $s_{12}^\ell\sim(M_\ell)_{12}/(M_\ell)_{22}$
and $s_{13}^\ell\sim(M_\ell)_{13}/(M_\ell)_{33}$. Here, we focus
on the neutrino mass matrix with a pseudo-Dirac structure. If there are no exact 
relations between different entries in $M_\nu$, then the most general structure 
that is consistent with $\epsonu\ll1$ and $s_{13}^\nu\ll1$ is
\eqn\genmnu{M_\nu=m\pmatrix{y_{11}&Y_{12}&Y_{13}\cr
Y_{12}&y_{22}&y_{23}\cr Y_{13}&y_{23}&Y_{33}\cr},}
where
\eqn\oomadel{Y_{12}\sim1,\ \ \ y_{ij}\ll1.}
As concerns $Y_{13}$ and $Y_{33}$, there are three different options:
\eqn\oomaz{\eqalign{
(i)&\ \ \ Y_{13}\lsim1,\ \ \ Y_{33}\gg1,\cr
(ii)&\ \ \ Y_{13}\lsim1,\ \ \ Y_{33}\equiv y_{33}\ll1,\cr
(iii)&\ \ \ Y_{13}\equiv y_{13}\ll1,\ \ \ Y_{33}\sim1.\cr}}
(Explicit examples of models in the literature that realize these options 
are presented in section 7.)
It is also convenient to define the matrix
\eqn\hatmnu{\hat M_\nu=R_{13}^T(\theta_{13}^\nu)R_{23}^T(\theta_{23}^\nu)
M_\nu R_{23}(\theta_{23}^\nu)R_{13}(\theta_{13}^\nu).}
By definition, it is block diagonal. The requirement that
$\epsonu\ll1$ restricts the form of the (12) block: 
\eqn\hamnu{\hat M_\nu=m\pmatrix{\delta_1&1&0\cr1&\delta_2&0\cr
0&0&Y_{3}\cr};\ \ \ \ |\delta_1|,\ |\delta_2|\ \ll1.}
Both $\Delta m^2_{21}/m^2$ and $\epsonu$ depend on
only $\delta_1$ and $\delta_2$:
\eqn\epnudel{\eqalign{
{\Delta m^2_{12}\over m^2}\ =&\ 2|\delta_1^*+\delta_2|,\cr
\epsonu\ =&\ {|\delta_2|^2-|\delta_1|^2\over2|\delta_1^*+\delta_2|}.\cr}}

We now present, for the three cases of eq. \oomaz, expressions for
$s_{23}^\nu$, $s_{13}^\nu$, $\delta_1$ and $\delta_2$ to first order in 
the small parameters $y_{ij}$. For case (i), we take $Y_{12}=1$ and obtain:
\eqn\largez{\eqalign{
s_{13}^\nu\ =&\ {Y_{13}/Y_{33}},\ \ \ 
s_{23}^\nu\ =\ {(s_{13}^\nu Y_{12}+y_{23})/Y_{33}},\cr
\delta_1\ =&\ y_{11}-{Y_{13}^2/Y_{33}},\ \ \ 
\delta_2\ =\ y_{22}.\cr}}
For case (ii), we take $c_{23}^\nu Y_{12}-s_{23}^\nu Y_{13}=1$ and obtain:
\eqn\smallz{\eqalign{
\tan\theta_{23}^\nu\ =&\ -{Y_{13}/Y_{12}},\ \ \ 
s_{13}^\nu\ =\ c^\nu_{23}s^\nu_{23}(y_{33}-y_{22})
-((c_{23}^\nu)^2-(s_{23}^\nu)^2)y_{23},\cr
\delta_1\ =&\ y_{11},\ \ \ 
\delta_2\ =\ (c_{23}^{\nu})^2y_{22}+(s_{23}^{\nu})^2y_{33}
-2s^\nu_{23}c^\nu_{23}y_{23}.\cr}}
For case (iii), we take $Y_{12}=1$ and obtain: 
\eqn\interz{\eqalign{
s_{23}^\nu\ =&\ {y_{23}Y_{33}+y_{13}Y_{12}\over Y_{33}^2-Y_{12}^2},\ \ \ 
s_{13}^\nu\ =\ {y_{13}Y_{33}+y_{23}Y_{12}\over Y_{33}^2-Y_{12}^2},\cr
\delta_1\ =&\ y_{11},\ \ \ 
\delta_2\ =\ y_{22}.\cr}}

We would like to emphasize several points related to the results
derived above (some of the statements below were previously made in ref.
\ref\DuJo{G. Dutta and A.S. Joshipura,
 Phys. Rev. D51 (1994) 3838,  hep-ph/9405291.}\
in the context of a specific class of textures for the Dirac and
Majorana mass matrices in the seesaw model):

(a) Eq. \epnudel\ implies that flavor models where $\epsonu$ gives the 
dominant contribution to $\epson$ can be strongly constrained by a
measurement of $\epson$. Since $\delta_1$ and $\delta_2$ depend on 
different entries of $M_\nu$, we expect no exact cancellations in their 
contribution to $\epsonu$. Consequently, one will be able to use the measured 
{\it size} of $\epson$ to estimate the size of the larger between $|\delta_1|$ 
and $|\delta_2|$, and the {\it sign} of $\epson$ to tell which is larger.

(b) Eq. \mnssma\ implies that observable deviations from maximal
mixing in vacuum oscillations, $1-\sin^22\theta_{12}=\epson^2\neq0$, 
can strongly constrain flavor models. For models with a small $s_{23}^\nu$, 
we have
\eqn\lepvsnu{\epson^2\sim{\rm max}\left({\delta_2^2\over4},
{\delta_1^2\over4},4(s_{12}^\ell)^2\right).}
If vacuum oscillations show an observable deviation from maximal
mixing, say, $\epson^2\sim0.1$, it would be difficult to explain
it with a parametrically suppressed $\delta_2$, $\delta_1$ and $s_{12}^\ell$.
The accidental factor of sixteen, however, between the $s_{12}^\ell$
and the $\delta_i$ contributions in eq. \lepvsnu\ favors $s_{12}^\ell$
as the major source for such a large effect.

(c) Eq. \epnudel\ reveals interesting relations between the mass hierarchy
and the mixing. The parameters $\epsonu$ and $\Delta m^2_{21}/m^2$
are of the same order of magnitude. Therefore, within models where
$\epsonu$ gives the dominant contribution to $\epson$, one will
be able to use the measured values of $\epson$ and $\Delta m^2_{21}$
to estimate the mass scale $m$ of the pseudo-Dirac neutrino pair. 
If the contributions to $\epson$ related to $s_{12}^\ell$ and/or to
$s_{13}^\ell$ are larger than the contribution related to $\epsonu$, then the
relation between $\epson$ and $\Delta m^2_{21}/m^2$ is lost and,
in particular, $\epson\gg\Delta m^2_{21}/m^2$ is possible.
In any case, if there are no exact relations between
entries of the lepton mass matrices, we expect 
\eqn\oomineq{|\epson|\gsim\Delta m^2_{21}/m^2.}
This relation can be used in two ways. First, measurements of $|\epson|$ and 
of $\Delta m^2_{21}$ would give a lower bound on $m$. Second, in our framework
we have $m^2\lsim \Delta m^2_{\rm atm}$ and therefore we expect
\eqn\ooman{|\epson|\gsim{\Delta m^2_{\rm sun}\over\Delta m^2_{\rm atm}}.}
This constraint is particularly powerful if the LMA solution (see eq. \SNdata)
is realized in Nature since then 
$\Delta m^2_{\rm sun}/\Delta m^2_{\rm atm}\sim0.01$.

%%%%%%%%%%%%%%%%%%%%%%%%%%%%%%%%%
%%%%%%%%%%%%%%%%%%%%%%%%%%%%%%%%%
\newsec{Radiative Corrections}
\nref\ChPl{P.H. Chankowski and Z. Pluciennik, Phys.
 Lett. B316 (1993) 312, hep-ph/9306333.}%
\nref\BLP{K.S. Babu, C.N. Leung and J. Pantaleone,
 Phys. Lett. B319 (1993) 191, hep-ph/9309223.}%  
\nref\ElLo{J. Ellis and S. Lola,
 Phys. Lett. B458 (1999) 310, hep-ph/9904279.}%
\nref\CEIN{J.A. Casas, J.R. Espinosa, A. Ibarra and I. Navarro,
 Nucl. Phys. B556 (1999) 3, hep-ph/9904395;
 Nucl. Phys. B569 (2000) 82, hep-ph/9905381;
 JHEP 9909 (1999) 015, hep-ph/9906281.}%
\nref\BRS{R. Barbieri, G.G. Ross and A. Strumia,
 JHEP 9910 (1999) 020, hep-ph/9906470.}%  
\nref\EMa{E. Ma, J. Phys. G25 (1999) L97, hep-ph/9907400 .}%  
\nref\HMOS{N. Haba, Y. Matsui, N. Okamura and M. Sugiura,
 Prog. Theor. Phys. 103 (2000) 145, hep-ph/9908429.}%  
\nref\CKP{P.H. Chankowski, W. Krolikowski and S. Pokorski,
 Phys. Lett. B473 (2000) 109, hep-ph/9910231.}%
\nref\BDMP{K.R.S. Balaji, A.S. Dighe, R.N. Mohapatra and M.K. Parida,
 Phys. Rev. Lett. 84 (2000) 5034, hep-ph/0001310.}% 
We consider the effect of radiative corrections on neutrino mass matrices 
which at a high energy scale $\Lambda$ have the pseudo-Dirac structure 
\genmnu. In particular, we ask whether at some low energy scale $\mu$ that is 
relevant to the solar neutrinos, a significant deviation from maximal mixing 
could be induced by renormalization group evolution (RGE).
We take as our framework the minimal supersymmetric Standard Model.
(Our results apply also to the Standard Model, but there the smallness
of the charged lepton Yukawa couplings guarantees that the radiative
corrections are negligible for our purposes.) 

The important parameter for our purposes is related to
the Yuakawa coupling of the tau lepton:
\eqn\defet{\epst\equiv-{g_\tau^2\over(4\pi)^2}(1+\tan^2\beta)
\ln{\Lambda\over\mu}.}
Here $g_\tau(1+\tan^2\beta)^{1/2}=m_\tau/\vev{\phi_d}$ is the tau Yukawa 
coupling in the supersymmetric standard model. The $\epst$ parameter
could be of ${\cal O}(0.01)$ for large $\tan\beta$. (Within the SM,
one has to replace $(1+\tan^2\beta)$ with $-1/2$, which gives
$\epst\sim10^{-6}$.) Define a matrix
\eqn\defIt{I_\tau={\rm diag}(1,1,1+\epst).}
We denote the neutrino mass scale at the high scale $\Lambda$ by 
$M^{\rm HE}_\nu$. Then, up to universal corrections and negligibly small 
effects of the muon and electron Yukawa couplings, the renormalized
neutrino mass matrix at a scale $\mu$ below $\Lambda$ is given
in logarithmic approximation by \refs{\ChPl-\BDMP}
\eqn\renMnu{M_\nu=I_\tau M_\nu^{\rm HE} I_\tau.}

In this section, the parameters that relate to $M_\nu^{\rm HE}$
and to its diagonalization are denoted, as before, by $s_{ij}^\nu$ and
$\delta_i$. They can be expressed in terms of the
entries of $M_\nu^{\rm HE}$ according to equations \largez, \smallz\
and \interz. In other words, we have
\eqn\formMHE{M_\nu^{\rm HE}=mR_{23}(\theta_{23})R_{13}(\theta_{13})
\pmatrix{\delta_1&1&0\cr 1&\delta_2&0\cr 0&0&Y_{3}\cr}
R_{13}^T(\theta_{13})R_{23}^T(\theta_{23}).}
The parameters that relate to $M_\nu$ and to its diagonalization
will be denoted by $\hat s_{ij}$ and $\hat\delta_i$. The difference between
them and the corresponding $s_{ij}$ and $\delta_i$ parameters
vanishes in the limit $\epst\rightarrow0$.
The main question that we would like to investigate is whether
the differences $\hat\delta_{1,2}-\delta_{1,2}$ are of ${\cal O}(\epst)$ 
or $\ll{\cal O}(\epst)$. In the latter case the radiative corrections
can be safely neglected.

After a cumbersome but straightforward calculation, we find
the following leading corrections:
\eqn\soldel{\eqalign{
\hat s^\nu_{23}-s^\nu_{23}\ =&\ {1+Y_3^2\over1-Y_3^2}(c_{23}^{\nu})^2s_{23}^\nu\epst
+{\cal O}(s^\nu_{13}\epst),\cr
\hat s^\nu_{13}-s^\nu_{13}\ =&\ {2Y_3\over1-Y_3^2}c_{23}^{\nu}s_{23}^\nu\epst
+{\cal O}(s^\nu_{13}\epst),\cr
\hat\delta_1-\delta_1\ =&\ {4Y_3^2\over Y_3^2-1}s^\nu_{13}c^\nu_{23}s^\nu_{23}\epst
+{\cal O}(\epst^2),\cr
\hat\delta_2-\delta_2\ =&\ -{4\over Y_3^2-1}s^\nu_{13}c^\nu_{23}s^\nu_{23}\epst
+{\cal O}(\epst^2).\cr}}
From the expressions for $\hat\delta_1$ and $\hat\delta_2$, we obtain:
\eqn\emasdif{\eqalign{
{\Delta m^2_{21}\over m^2}\ =&\ 
2|\delta_1+\delta_2+4s^\nu_{13}c_{23}^{\nu}s_{23}^\nu\epst|,\cr
\hat\epsonu\ =&\ 
{(\delta_2-\delta_1)\over2}\ {\delta_2+\delta_1+4s^\nu_{13}c^\nu_{23}s^\nu_{23}\epst
\over|\delta_1+\delta_2+4s^\nu_{13}c^\nu_{23}s^\nu_{23}\epst|}
+2{1+Y_3^2\over1-Y_3^2}{(\delta_2+\delta_1)s^\nu_{13}c^\nu_{23}s^\nu_{23}\epst
\over|\delta_1+\delta_2+4s^\nu_{13}c^\nu_{23}s^\nu_{23}\epst|}.\cr}}
Eq. \emasdif\ should be compared to eq. \epnudel.

\nref\CHOOZ{CHOOZ collaboration, M. Apollonio {\it et al.},
 Phys. Lett. B420 (1998) 397, hep-ex/9711002;
 Phys. Lett. B466 (1999) 415, hep-ex/9907037.}%
\nref\SKatm{Super-Kamiokande Collaboration, Y. Fukuda {\it et al.},
 Phys. Rev. Lett. 81 (1998) 1562, hep-ex/9807003.}%  
\nref\BWW{V. Barger, T.J. Weiler and K. Whishnant,
 Phys. Lett. B440 (1998) 1, hep-ph/9807319.}%
\nref\FLMS{G.L. Fogli, E. Lisi, A. Maronne and G. Scioscia,
 Phys. Rev. D59 (1999) 033001, hep-ph/9808205.}%
We would like to emphasize the following points concerning equations
\soldel\ and \emasdif:
\item{(a)} The change in $s^\nu_{23}$ is small, 
$(\hat s^\nu_{23}-s^\nu_{23})/s^\nu_{23}={\cal O}(\epst)$.
\item{(b)} The difference $\hat s^\nu_{13}-s^\nu_{13}$ is suppressed beyond the 
naive estimate of $\epst$: for large (small) $Y_3$ it is further suppressed
by $1/Y_3$ ($Y_3$) while for $Y_3\sim1$ it is suppressed by the small
$s^\nu_{23}$. Effectively then we have 
$(\hat s^\nu_{13}-s^\nu_{13})/s^\nu_{13}={\cal O}(\epst)$. 
\item{(c)} Our main result is that the RGE-induced deviation from maximal 
mixing and mass splitting are suppressed by $\epst s_{13}$. 
A combination of the CHOOZ result \CHOOZ\ and the SuperKamiokande results on 
atmospheric neutrinos \SKatm\ implies that $s_{13}$ is small \refs{\BWW,\FLMS}.
Consequently, the $\epst s_{13}$ suppression factor is constrained to be
below ${\cal O}(10^{-3})$. In the limit $s_{13}=0$, the leading effects are of 
order $\epst^2$ \BRS.  

To summarize the results of this section: We find that the contribution from 
radiative corrections to the deviation from maximal mixing is suppressed beyond 
the smallness of $\epst$. The leading corrections to $\epsonu$, $s_{13}^\nu$
and $\Delta m^2_{21}/m^2$ are ${\cal O}[\epst\times{\rm max}(s_{13},\epst)]$.
Model independently, the size of the effect is not larger than ${\cal O}(10^{-3})$.
(This correction could be important for the mass splitting in the
VAC$_{\rm L}$ solution.)
For the deviation from maximal mixing, this correction is too small to be observed.

%%%%%%%%%%%%%%%%%%%%%%%%%%%%%%%%%
%%%%%%%%%%%%%%%%%%%%%%%%%%%%%%%%%
\newsec{Non-Canonical Kinetic Terms}
Models with horizontal symmetries predict the structure of the
mass matrices in the basis where the horizontal charges are
well defined. This preferred interaction basis can, in general, be different
from the basis where the kinetic terms are canonically normalized 
\nref\DPS{E. Dudas, S. Pokorski and C.A. Savoy,
 Phys. Lett. B356 (1995) 45, hep-ph/9504292.}%
\nref\EyNi{G. Eyal and Y. Nir,
 Nucl. Phys. B528 (1998) 21, hep-ph/9801411.}% 
\refs{\LNSb,\DPS,\EyNi}. In particular, when heavy degrees of freedom
related to flavor physics are integrated out, the kinetic terms for the 
left-handed lepton doublets $L_i$ ($i=1,2,3$) can be modified to
\eqn\norkin{R_{ij}L_i^\dagger \gamma^\mu\partial_\mu L_j,}
where $R$ is a hermitian matrix. By rescaling of the
$L_i$ fields we can bring the diagonal entries of $R$ to equal unity,
\eqn\Rentries{R_{ii}=1,\ \ \ |R_{ij}|\ll1\ \ (i\neq j).}
One can always find a hermitian matrix $K$ that brings the kinetic
terms back to canonical normalization \LNSb:
\eqn\roleK{K^\dagger R K={\rm diag}(1,1,1).}
If the mass matrix in the basis where the kinetic terms
are of the non-canonical form \norkin\ is $M_\nu^{\rm NC}$, 
then the true mass matrix, that is the matrix in the basis with 
canonical kinetic terms, is given by
\eqn\defMnc{M_\nu=K^T M_\nu^{\rm NC} K.}
$K$ has the form
\eqn\formK{K=\pmatrix{1&k_{12}&k_{13}\cr k_{12}^*&1&k_{23}\cr
k_{13}^*&k_{23}^*&1\cr}.}
For simplicity, we again neglect CP violation and take $R$ and,
consequently $K$, to be real. 

We are interested in finding the effects of $k_{ij}\neq0$
on the deviation from maximal mixing and on the mass splitting. 
Our analysis follows similar lines to our study of radiative
corrections in the previous section. We take
\eqn\formMNC{M_\nu^{\rm NC}=mR_{23}(\theta^\nu_{23})R_{13}(\theta^\nu_{13})
\pmatrix{\delta_1&1&0\cr 1&\delta_2&0\cr 0&0&Y_3\cr}
R_{13}^T(\theta^\nu_{13})R_{23}^T(\theta^\nu_{23}).}
The parameters that relate to the matrix $M_\nu$ of eq. \defMnc\
are denoted by $\hat s^\nu_{ij}$ and $\hat\delta_i$.

For the differences between $\hat s^\nu_{ij},\ \hat\delta_i$ and the 
corresponding $s^\nu_{ij},\ \delta_i$, we find:
\eqn\solvehade{\eqalign{
\hat s^\nu_{23}-s^\nu_{23}\ =&\ {c^\nu_{23}\over Y_3^2-1}\left[(Y_3^2+1)
k_{23}((c^\nu_{23})^2-(s^\nu_{23})^2)+2Y_3(k_{12}s^\nu_{23}+k_{13}c^\nu_{23})
\right],\cr
\hat s^\nu_{13}-s^\nu_{13}\ =&\ {1\over Y_3^2-1}\left[(Y_3^2+1)
(k_{12}s^\nu_{23}+k_{13}c^\nu_{23})+2Y_3k_{23}((c^\nu_{23})^2-(s^\nu_{23})^2)
\right],\cr
\hat\delta_{2,1}-\delta_{2,1}\ =&\ 2(k_{12}c^\nu_{23}-k_{13}s^\nu_{23}).\cr}}
For the mass difference and deviation from maximal mixing, we obtain
\eqn\kmasdif{\eqalign{{\Delta m^2_{21}\over m^2}\ =&\ 
2|\delta_1+\delta_2+4(k_{12}c^\nu_{23}-k_{13}s^\nu_{23})|,\cr
\hat\epsonu\ =&\ {(\delta_2-\delta_1)\over2}\ {\delta_1+\delta_2
+4(k_{12}c^\nu_{23}-k_{13}s^\nu_{23})
\over|\delta_1+\delta_2+4(k_{12}c^\nu_{23}-k_{13}s^\nu_{23})|}.\cr}}
Terms of order $s^\nu_{13}k_{ij}$ contribute with different signs to
$\hat\delta_1$ and $\hat\delta_2$ and modify $\hat\epsonu$ in a qualitatively
different way. Quantitatively, however, these effects are negligible.

Before we analyze the consequences eqs. \solvehade\ and \kmasdif,
we would like to make two comments regarding the size of $k_{ij}$
in Froggett-Nielsen type models:
\item{1.} In most models of horizontal symmetries, we have
\eqn\smallk{k_{ij}\lsim s^\ell_{ij}.}
\item{2.} If two fields carry the same horizontal quantum numbers, $H(L_i)
=H(L_j)$, we can always define these fields in such a way that $k_{ij}=0$.
 
We would like to emphasize the following points:
\item{(a)} The changes in $s^\nu_{23}$ and $s^\nu_{13}$ are of ${\cal O}(k_{ij})$.
The effect can be significant for a small mixing angle. In particular, in
the supersymmetric framework, if $s_{ij}^\nu$ vanishes because of holomorphy
\LNSb, we expect such zeros to be lifted by these corrections.
The mixing angle is, however, still parametrically suppressed.
\item{(b)} The leading effect on $\epsonu$ does not change its size
but, if $4(k_{12}c^\nu_{23}-k_{13}s^\nu_{23})$ is not much smaller than
max($\delta_1,\delta_2$), can affect its sign. (With CP violating phases,
also the size of $\epsonu$ is affected, but in the Froggatt-Nielsen class of 
models the parametric suppression remains the same.)
\item{(c)} In principle, the $k_{ij}$-related corrections could enhance
$\Delta m^2_{21}/m^2$ compared to $\epsonu$ and therefore avoid \oomineq.
However, in models where the constraint \smallk\ holds, \oomineq\ is valid.

We conclude that, in general,  in models where the kinetic terms 
are normalized according to \norkin, sign($\epson$)  
does not give a useful constraint.

%%%%%%%%%%%%%%%%%%%%%%
%%%%%%%%%%%%%%%%%%%%%%
\newsec{An Effective Two Generation Framework}
In previous sections we took all the parameters in the Lagrangian to be
real. To understand some of the effects of phases, we analyze a
two generation model allowing for the most general phase structure.
          
We parametrize the two generation mixing matrix by
\eqn\parnbs{V=\pmatrix{c&se^{i\beta}\cr -s&ce^{i\beta}\cr},}
where $c\equiv\cos\theta_{12}$, $s\equiv\sin\theta_{12}$ and the
phase $\beta$ is physical but does not play a role in oscillation
experiments. We parametrize the diagonalizing matrices 
$V_\ell$ and $V_\nu$ in the following way:
\eqn\parV{V_\ell=\pmatrix{c_\ell&s_\ell e^{i\beta_\ell}\cr 
-s_\ell& c_\ell e^{i\beta_\ell}\cr},\ \ \ \ 
V_\nu=\pmatrix{c_\nu&s_\nu e^{i\beta_\nu}\cr 
-s_\nu& c_\nu e^{i\beta_\nu}\cr}.}
Using \VVVP, we can express the size of the mixing angle in terms of the
four parameters $s_\nu$, $s_\ell$, $\beta_\nu$ and $\beta_\ell$:
\eqn\findssq{s^2=c_\ell^2 s_\nu^2+s_\ell^2 c_\nu^2-
2\Re(c_\ell s_\ell c_\nu s_\nu e^{i(\beta_\ell-\beta_\nu)}).} 

The charged lepton mass matrix can be written as
\eqn\defMell{M_\ell=\pmatrix{m_{11}&m_{12}\cr m_{21}&m_{22}\cr}.}
Our assumption that $s_\ell$ is small requires that a certain
combination of entries is small:
\eqn\dellsma{|s_\ell|\simeq|\delta_\ell|\ll1,}
where
\eqn\defdell{\delta_\ell\equiv{m_{11}m_{21}^*+m_{12}m_{22}^*\over 
|m_{21}|^2+|m_{22}|^2-|m_{12}|^2-|m_{11}|^2}.}

The neutrino mass matrix is given by
\eqn\Mnuteff{M_\nu=m\pmatrix{\delta_1&1\cr 1&\delta_2\cr},\ \ \ 
|\delta_i|\ll1.}
We include the effects of radiative corrections, parametrized by
\eqn\defImu{I_\mu={\rm diag}(1,1+\epsilon_\mu);\ \ \ \ 
\epsilon_\mu\equiv-{g_\mu^2\over(4\pi)^2}(1+\tan^2\beta)\ln{\Lambda\over\mu},}
and of non-canonical kinetic terms, parametrized by
\eqn\defKtwo{K=\pmatrix{1&k\cr k^*&1\cr}.}
We find:
\eqn\truemnu{\eqalign{
\hat\delta_1\ =&\ \delta_1+2k^*,\cr
\hat\delta_2\ =&\ \delta_2+2k.\cr}}

We can now express the mass splitting $\Delta m^2/m^2$ and the
deviation from maximal mixing $\epson$ in terms of the three
parameters $\delta_\ell$ of eq. \defdell\ and $\hat\delta_1$ 
and $\hat\delta_2$ of eq. \truemnu:
\eqn\twomasdif{\eqalign{
{\Delta m^2\over m^2}\ =&\ 2|\hat\delta_1^*+\hat\delta_2|,\cr
\epson\ =&\ {|\hat\delta_2|^2-|\hat\delta_1|^2
\over2|\hat\delta_1^*+\hat\delta_2|}+2\Re\left[\delta_\ell\left(
{|\hat\delta_1^*+\hat\delta_2|\over\hat\delta_1^*+\hat\delta_2}
\right)\right].\cr}}

Eq. \twomasdif\ allows us to make (or to re-emphasize) the following points:

1. We again observe the accidental factor of four between the 
$\delta_\ell$ contribution and the $\hat\delta_i$ contribution to
$\epson$. A large measured value of $\epson$ might be a hint then
to the size of $\delta_\ell$.
 
2. The usefulness of an experimental determination of sign($\epson$) 
depends on the relative size of the  small parameters. If 
$|\delta_1|,|\delta_2|\gg|\dell|,|k|$, then sign($\epson$) depends on the 
relative {\it size} of $|\delta_1|$ and $|\delta_2|$ which is predicted by 
the models and a useful constraint can be derived.  On the other hand, 
if $|\delta_\ell|$ and/or $|k|$ are not smaller than both $|\delta_1|$
and $|\delta_2|$, then sign($\epson$) depends on the relative {\it phases} between 
$\delta_\ell$ or $k$ and $(\delta_1^*+\delta_2)$. Since generic models of approximate 
horizontal symmetries do not predict the phases, we cannot derive any useful 
constraint. 

%%%%%%%%%%%%%%%%% 
%%%%%%%%%%%%%%%%%
\newsec{Abelian horizontal symmetries}
The most natural application of our results is in the framework
of approximate Abelian horizontal symmetries.
To understand the principles of this framework, let us take
the simplest example of a horizontal symmetry, $H=U(1)$, that is
broken by a single small parameter. We denote the breaking
parameter by $\lambda$ and assign to it a horizontal charge $-1$.
Wherever numerical values are relevant, we take $\lambda=0.2$
(so that it is of the order of the Cabibbo angle).
Within a supersymmetric framework, the following selection rules apply:
\item{a.} Terms in the superpotential that carry an integer $H$-charge
$n\geq0$ are suppressed by $\lambda^n$.  Terms with $n<0$ vanish by
holomorphy.
\item{b.} Terms in the K\"ahler potential that carry an integer 
$H$-charge $n$ are suppressed by $\lambda^{|n|}$.

We are particularly interested in the leptonic Yukawa terms:
\eqn\Lyuk{-{\cal L}_{Y}=Y^\ell_{ij}L_i\bar\ell_j\phi_d+{Y^\nu_{ij}\over M}
L_iL_j\phi_u\phi_u+{\rm h.c.},}
where $i=1,2,3$ is a generation index,
$L_i$ are lepton doublet fields, $\bar\ell_j$ are lepton
charged singlet fields, and $\phi_u$ and $\phi_d$ are the two
Higgs fields. The couplings $Y_{ij}$ are dimensionless Yukawa
couplings and $M$ is a high energy scale. The Yukawa terms come
from the superpotential. If the sum of the horizontal charges
in a particular term is a positive integer, then the resulting
mass term is suppressed as follows:
\eqn\selrul{\eqalign{
(M_\ell)_{ij}\ \sim&\ \vev{\phi_d}\lambda^{H(L_i)+H(\bar\ell_j)+H(\phi_d)},\cr
(M_\nu)_{ij}\ \sim&\ {\vev{\phi_u}^2\over M}\lambda^{H(L_i)+H(L_j)+2H(\phi_u)}.\cr}} 
Otherwise, {\it i.e.} if the sum of charges is negative or non-integer,
the Yukawa coupling vanishes. We use the $\sim$ sign to emphasize that
there is an unknown, independent, order one coefficient for each term (except
for the relation $(M_\nu)_{ij}=(M_\nu)_{ji}$)).

To understand the possible implications of close-to-maximal mixing
on theoretical model building, we imagine that  future measurements 
will give
\eqn\largesel{\epson\sim\lambda.}
We examine the consequences of such a constraint on three classes of models 
in the literature. We find that two classes of models will be excluded, 
while in the other a unique model is singled out that is consistent
with all the requirements.

%%%%%%%%%%%%%%%%%%%%%%%%
\subsec{Holomorphic zeros}
Option (i) of eq. \oomaz\ has been realized in
the framework of supersymmetric Abelian horizontal symmetries, 
where holomorphic zeros can induce a large 23 mixing
together with large 23 mass hierarchy
\ref\GNS{Y. Grossman, Y. Nir and Y. Shadmi,
 JHEP 9810 (1998) 007, hep-ph/9808355.}.
The horizontal symmetry is $U(1)_1\times U(1)_2$ with breaking parameters
\eqn\brepar{\lambda_1(-1,0),\ \ \ \lambda_2(0,-1);\ \ \ 
\lambda_1\sim\lambda_2\sim\lambda=0.2.}

We impose four requirements on the model: Large 23 mixing, $s_{23}\sim1$;
Large hierarchy, $m_2/m_3\ll1$; $\nu_1$ and $\nu_2$ form a pseudo-Dirac neutrino,
$\Delta m^2_{12}\ll m^2$;  A deviation from maximal mixing given by 
$\epson\sim\lambda$
(this is the {\it hypothetical} constraint from solar neutrino measurements).
We find that there is a single set of horizontal charge assignments
to the Higgs and lepton doublets that is consistent with all four requirements:
\eqn\setcha{\phi_u(0,0),\ \ \ \phi_d(0,0),\ \ \ L_1(1,0),\ \ \ 
L_2(-1,1),\ \ \ L_3(0,0).}
(The choice is single up to trivial shifts by hypercharge which is
an exact symmetry of the model, by a Peccei-Quinn symmetry that is
an accidental symmetry of the Yukawa sector, and by lepton number
if it only changes the overall neutrino mass scale and can be absorbed
in the parameter $M$, and up to trivial exchange of $U(1)_1\leftrightarrow 
U(1)_2$.) We find then a unique structure for $M_\nu$:
\eqn\Mnuthree{M_\nu\sim {\vev{\phi_u}^2\over M}\pmatrix{\lambda^2&\lambda&
\lambda\cr \lambda&0&0\cr \lambda&0&1\cr}.}
This matrix is of the form \genmnu\ with option (i) of eq. \oomaz.
Therefore, eqs. \largez\ can be applied. To have $s_{23}\sim1$ and 
large enough $\epson$, together with acceptable charged lepton mass hierarchy,
we can choose, for example,
\eqn\chacha{\bar\ell_1(3,4),\ \ \ \bar\ell_2(3,2),\ \ \ \bar\ell_3(3,0),}
which gives
\eqn\chamas{M_\ell\sim\vev{\phi_d}\pmatrix{\lambda^8&\lambda^6&\lambda^4\cr
\lambda^7&\lambda^5&\lambda^3\cr \lambda^7&\lambda^5&\lambda^3\cr}.}
The parametric suppression of the physical parameters is then as follows:
\eqn\lepmas{m_\tau/\vev{\phi_d}\sim\lambda^3,\ \ \ m_\mu/m_\tau\sim\lambda^2,
\ \ \ m_e/m_\mu\sim\lambda^3,}
\eqn\nepars{{\Delta m^2_{21}/\Delta m^2_{23}}\sim\lambda^3,\ \ 
\Delta m^2_{12}/m^2\sim\lambda,}
\eqn\mixing{s_{23}\sim1,\ \ \ s_{13}\sim\lambda,\ \ \ \epson\sim\lambda.}

The corrections due to a non-canonical kinetic terms,
\eqn\kwithL{k_{23}\sim\lambda^2,\ \ \ 
k_{12}\sim\lambda^3,\ \ \ k_{13}\sim\lambda,}
leave eqs. \nepars\ and \mixing\ unchanged.

Within the framework of Abelian horizontal symmetries, it is particularly
interesting to find predictions for relations among the physical parameters
that are independent of a specific choice of horizontal charges.
In the quark sector, there is a single such relation \LNSb,
$|V_{us}|\sim|V_{ub}/V_{cb}|$. In the lepton sector, when singlet
neutrinos play no role, there are three such relations 
\ref\GrNiL{Y. Grossman and Y. Nir,
 Nucl. Phys. B448 (1994) 30, hep-ph/9502418.}. 
For the class of models where holomorphic zeros give a pseudo-Dirac
structure in the 12 sector but do not affect the parameters that are
related to the third generation (the model presented in this
subsection belongs to this class), we have the following relations:
\eqn\chaind{\eqalign{
\epson\ \sim&\ s_{13}/s_{23},\cr
m/m_3\ \sim&\ s_{13}s_{23},\cr
\Delta m^2_{12}/m^2\ \sim&\ s_{13}/s_{23}.\cr}}
The first of these relations, which involves only mixing angles,
can be tested if oscillation experiments measure $\epson$ and $s_{13}$.
The last two relations can be combined to give another testable relation.
\eqn\miatsu{\Delta m^2_{\rm sun}/\Delta m^2_{\rm atm}\sim s_{13}^3 s_{23}.}

%%%%%%%%%%%%%%%%%%%%%%%%%%%%%%%%%%%%
\subsec{$L_e-L_\mu-L_\tau$ symmetry}
Option (ii) of eq. \oomaz\ can be realized in a particulary interesting 
frameowork of approximate $L_e-L_\mu-L_\tau$ symmetry
\nref\BHSSW{R. Barbieri, L.J. Hall, D. Smith, A. Strumia
and N. Weiner, JHEP 9812 (1998) 017, hep-ph/9807235.}%
\nref\JoRi{A.S. Joshipura and S.D. Rindani, hep-ph/9811252.}%
\nref\FrGl{P.H. Frampton and S.L. Glashow,
 Phys. Lett. B461 (1999) 95, hep-ph/9906375.}%
\nref\MPP{R.N. Mohapatra, A. Perez-Lorenzana and C.A. de S. Pires,
 Phys. Lett. B474 (2000) 355, hep-ph/9911395.}%
\nref\ChKe{K. Cheung and O.C.W. Kong,
 Phys. Rev. D61 (2000) 113012, hep-ph/9912238.}% 
\nref\ShTa{Q. Shafi and Z. Tavartkiladze,
 Phys. Lett. B482 (2000) 145, hep-ph/0002150.}% 
\refs{\BHSSW-\ShTa}. The symmetry is broken by small parameters, 
$\epsp$ and $\epsm$ of charges $+2$ and $-2$, respectively
\BHSSW. The neutrino mass matrix has the following form:
\eqn\Mnu{M_\nu\sim {\vev{\phi_u}^2\over M}\pmatrix{\epsm&1&1\cr
1&\epsp&\epsp\cr 1&\epsp&\epsp\cr}.}
This matrix is of the form \genmnu\ with option (ii) of eq. \oomaz.
Therefore, eqs. \smallz\ can be applied. We find:
\eqn\masses{m_{1,2}=m\left(1\pm{\cal O}[{\rm max}(\epsp,\epsm)]
\right),\ \ \ m_3=m{\cal O}(\epsp),}
\eqn\angles{s_{23}^\nu={\cal O}(1),\ \ \ s_{13}^\nu={\cal O}(\epsp),\ \ \ 
\epsonu={\cal O}[{\max}(\epsp,\epsm)].}

The charged lepton mass matrix has the form \BHSSW:
\eqn\Mell{M_\ell\sim \vev{\phi_d}\pmatrix{
\lambda_e&\lambda_\mu\epsm&\lambda_\tau\epsm\cr
\lambda_e\epsp&\lambda_\mu&\lambda_\tau\cr
\lambda_e\epsp&\lambda_\mu&\lambda_\tau\cr},}
where the $\lambda_i$ allow for a generic approximate symmetry that
acts on the SU(2)-singlet charged leptons. Such a symmetry, however,
does not affect the relevant diagonalizing angles:
\eqn\angell{s_{23}^\ell={\cal O}(1),\ \ \ s_{13}^\ell={\cal O}(\epsm),\ \ \ 
s_{12}^\ell={\cal O}(\epsm).}
Eqs. \angles\ and \angell\ lead to the following estimates of the
physical mixing angles:
\eqn\angphy{
s_{23}={\cal O}(1),\ \ \  
s_{13}={\cal O}[{\rm max}(\epsp,\epsm)],\ \ \ 
\epson={\cal O}[{\max}(\epsp,\epsm)].}

We can also estimate the corrections due to non-canonical kinetic terms:
\eqn\kwithL{k_{23}=0,\ \ \ 
k_{12},k_{13}={\cal O}[{\max}(\epsp,\epsm)].}
This leaves the parametric suppression of the physical parameters unchanged.

From eqs. \masses\ and \angphy\ we obtain: 
\eqn\spcial{\epson={\cal O}(\Delta m^2_{\rm sun}/\Delta m^2_{\rm atm}).}
Measurements of $\Delta m^2_{ij}$ and of $\epson$ can then lead to the exclusion 
of this model \BHSSW. For example, if $\Delta m^2_{\rm sun}/\Delta m^2_{\rm atm}\leq
10^{-2}$ and $\epson\geq0.1$ are established, the model will be excluded.

%%%%%%%%%%%%%%%%%%%
\subsec{Models with two breaking parameters}
Option (iii) of eq. \oomaz, that is hierarchy of mass splittings without
hierarchy of masses, has been realized in the framework of non-anomalous
horizontal $U(1)_H$ symmetry
\ref\NiSh{Y. Nir and Y. Shadmi, JHEP 9905 (1999) 023, hep-ph/9902293.}.
The symmetry is broken by two small parameters of opposite charges and
equal magnitudes:
\eqn\brenan{H(\lambda)=+1,\ \ \ H(\bar\lambda)=-1;\ \ \ 
\lambda=\bar\lambda\sim0.2.}
Then, the following selection rule applies: terms in the superpotential
or in the Kahler potential that carry an (integer) $H$-charge $n$ are suppressed
by $\lambda^{|n|}$. 
The three neutrino masses are of the same order of magnitude, but the
mass splitting between $\nu_1$ and $\nu_2$ is small if we have
\eqn\simmas{\eqalign{
|H(L_1)+H(L_2)|\ =&\ 2|H(L_3)|,\cr
|H(L_1)+H(L_2)|\ <&\ 2|H(L_1)|,\ 2|H(L_2)|.\cr}}

From eq. \interz\ we learn that
\eqn\epnuna{\epsonu\sim{\rm max}\left(\lambda^{2|H(L_1)|-|H(L_1)+H(L_2)|},
\lambda^{2|H(L_2)|-|H(L_1)+H(L_2)|}\right).}
A typical contribution to $s_{12}^\ell$ is given by
\eqn\sonna{s_{12}^\ell\sim\lambda^{|H(L_1)+H(\bar\ell_2)|-
|H(L_2)+H(\bar\ell_2)|}.}
The important point here is that the first condition in eq. \simmas\
requires that $H(L_1)$ and $H(L_2)$ are either both even or both odd.
Eqs. \epnuna\ and \sonna\ give then an upper bound on $\epson$,
\eqn\epsna{\epson\lsim\lambda^2.}
We conclude that if experiments find $\epson\sim\lambda$, this type
of models will be strongly disfavored.

%%%%%%%%%%%%%%%%%%%
\subsec{Alignment}
We would like to make a comment on a particular class
of supersymmetric models, where there is no degeneracy among
the sleptons and the only mechanism to suppress the supersymmetric
contributions to lepton flavor changing decays is {\it alignment}
\nref\NiSe{Y. Nir and N. Seiberg,
 Phys. Lett. B309 (1993) 337, hep-ph/9304307.}%
\refs{\NiSe,\LNSb,\GrNiL}, that is small mixing angles in the 
neutralino-lepton-slepton couplings. In such models, there is a 
strong constraint on $s_{12}^\ell$ (see {\it e.g.}
\ref\FNS{J. Feng, Y. Nir and Y. Shadmi,
 Phys. Rev. D61 (2000) 113005, hep-ph/9911370.}):
\eqn\muegam{{B(\mu\rightarrow e\gamma)\over1.2\times10^{-11}}
\sim\left({s_{12}^\ell\over2\times10^{-3}}\right)^2\left({
100\ GeV\over m(\tilde\ell)}\right)^4<1,}
where $m(\tilde\ell)$ is the average slepton mass.
In these models it is then particularly difficult to explain
a large deviation from maximal mixing. If the dominant source
of deviation from maximal mixing is $s_{12}^\ell$, we have
\eqn\maxeps{\epson\simeq2s_{12}^\ell\lsim4\times10^{-3}
\left({m(\tilde\ell)\over 100\ GeV}\right)^2.}

%%%%%%%%%%%%%%%%%%%%%%%%%%%%%
%%%%%%%%%%%%%%%%%%%%%%%%%%%%%
\newsec{Conclusions}
If the solar neutrino problem is solved by a large mixing angle solution,
and if the mixing is established to be close to maximal but not
precisely maximal, then interesting constraints for theoretical model
building would arise. Specifically, experiments may measure the size and
the sign of the small parameter $\epson$ defined by
\eqn\coneps{\sin^2\theta_{12}\equiv{1\over2}(1-\epson).}
Flavor models can account for a small $\epson$ by forcing a pseudo-Dirac
structure on the neutrino mass matrix through an approximate horizontal
symmetry,
\eqn\conmnu{M_\nu^{(2)}\sim m\pmatrix{\delta_1&1\cr 1&\delta_2\cr},\ \ \ 
|\delta_1|,|\delta_2|\ll1.}

We focus on models where there are no exact relations between
different entries of the lepton mass matrices (except for
$(M_\nu)_{ij}=(M_\nu)_{ji}$). Our main points are the following:

1. The most powerful constraints would arise if $\delta_1$ and/or $\delta_2$ are 
the dominant sources of $\epson$. Then the size of $|\epson|$ gives the size
of the larger between $|\delta_1|$ and $|\delta_2|$ while the sign of $\epson$
determines which of the two is larger. Moreover, the mass scale of the solar
neutrinos (and not only their mass-squared splitting) can be estimated,
$m^2\sim\Delta m^2_{21}/|\epson|$.

2. If the dominant source of $\epson$ is a small angle in the diagonalizing
matrix for the charged lepton mass matrix, $s_{12}^\ell$, then $|\epson|$
constrains the size of $s_{12}^\ell$ but sign($\epson$) is unlikely to test
the theoretical models. The order of magnitude relation between $|\epson|$
and $\Delta m^2_{12}/m^2$ is lost, but there is still a useful inequality,
$|\epson|\gsim\Delta m^2_{21}/m$. 

3. Radiative corrections do not play a significant role in $\epson$
and in $s_{13}$. They are supppressed by the tau Yukawa coupling, 
by a loop factor and by $s_{13}$. Consequently, their effect is below 
the level of $10^{-3}$.

4. In models of horizontal symmetries where the kinetic terms
are not canonically normalized, sign($\epson$) depends on
the kinetic terms as well and is unlikely to test the models. 

It remains to be seen whether future developments in solar neutrino
experiments would make a convincing case for the intriguing scenario
of pseudo-Dirac neutrinos
\ref\GNPS{M.C. Gonzalez-Garcia, Y. Nir, C. Pena-Garay and A. Smirnov,
to appear.}.

\vskip 1.0cm

\centerline{\bf Acknowledgments}
I thank John Bahcall, Plamen Krastev and Alexei Smirnov for useful
discussions and correspondence.
Partial support to this work was provided by the Department of Energy 
under contract No.~DE--FG02--90ER40542, by the Ambrose Monell Foundation, 
by AMIAS (Association of Members of the Institute for Advanced Study),  
by the Israel Science Foundation founded by the Israel Academy of Sciences
and Humanities, and by the Minerva Foundation (Munich).  

\listrefs
\end